\documentclass[10pt]{article}
\usepackage{amsfonts}
\usepackage{amsmath}
\title{On Gravitational Radiation by a Quantum Bound System}
\author{A. Jahan\\Research Institute for Astronomy and Astrophysics of Maragha
(RIAAM)\\ Maragha, IRAN, P. O. Box: 55134 - 441\\jahan@riaam.ac.ir}
\date{}
\usepackage[english]{babel}
\begin{document}
\maketitle
\begin{abstract}
A method based on the path integral approach is engaged to consider the gravitational emission from a quantum mechanical bound system in a locally inertial frame. In such a frame, interaction between the electromagnetic (bound potential) and gravitational fields can be neglected resulting in the less mathematical complexity. The final outcome is in agreement with the previous result for the radiation intensity of emitted gravitons due to decay of bound states in TT gauge.
\end{abstract}
\section{Introduction}
There is much literature on the interaction of gravitational field with non-relativistic quantum mechanical systems. Some examples include the Hydrogen atom [1-7], the many-body systems and harmonic oscillator [7, 9], and the quantized lattice vibrations of solids [10-14] . More recent considerations on the graviton emission and absorbtion by the atomic levels, reveal the significant role of locally inertial frame  in studying such systems: a locally inertial frame is the most relevant reference frame to overcome the difficulty associated with the gauge invariance problem of the interaction Lagrangian of systems with gravitational field [5, 6]. It is pointed out that some discrepancies for the graviton emission and absorbtion cross sections of the atomic levels, spread over the literature, originates from the gauge invariance violation of the interaction term in TT (transverse-traceless) gauge. However, it is demonstrated that such a problem could be avoided by moving on a locally inertial frame. As a result, there is a modification in the energy-momentum tensor and the contribution due to the interaction between the gravitational and electromagnetic fields becomes negligible [5, 6]. \\
In this note, motivated by the considerations in [10], we apply the path integral formalism to derive the spontaneous graviton emission rate by a quantum bound system in a locally inertial frame. Such an approach was originally engaged in context of QED to deal with the spontaneous photon emission by the atomic levels [15, 16]. We briefly review the notion of locally inertial frame in next section. Then, in section 3, we introduce the effective action gained by integrating out the gravitational field. In section 4, we derive the decay rate of a given energy level to the levels with lower energies due to the spontaneous graviton emission. The Greek and Roman indices for the space-time and spatial coordinates respectively are used. We assume $\hbar=c=1$.
\section{Locally Inertial Frame}
The interaction Hamiltonian density of matter with the gravitational field is given by
\begin{equation}\label{1}
\mathcal H_{int}=\frac{1}{2}h_{\mu\nu}T^{\mu\nu}.
\end{equation}
where $T_{\mu\nu}$ is the energy-momentum tensor of matter, which we suppose to be a system of particles bounded by an inter-particle potential, say Coulomb potential. In TT gauge, the Hamiltonian density (1) for a quantum mechanical system modifies to
\begin{equation}\label{1}
H_{int}=-\frac{1}{2m}h^{ij}\partial_i\partial_j+H'.
\end{equation}
where $H'$ represents the interaction between the gravitation field and bound potential. A rather sophisticated way of obtaining (2) is to use the Wheeler-De Witt equation [7, 17]. The above expression for the interaction Hamiltonian is the common starting point in most considerations cited in section 1. The problem with Hamiltonian (1) is that it is not invariant under the infinitesimal transformation of the coordinates and to solve this trouble one has to consider a reference frame which is locally inertial [5, 6]. In a locally inertial frame, which is almost Minkowskian, the dominant term of interaction Hamiltonian is
\begin{equation}\label{1}
H_{int}\approx\frac{1}{2} mh_{00}.
\end{equation}
and one can neglect the interaction between the graviton and the field causing the bound potential, setting $H'= 0$ [5, 6]. Now we impose the TT gauge condition by demanding $h_{0\mu}=h^\mu_{\;\mu}=0$ as a subset of the harmonic gauge $\partial^\alpha h_{\alpha\beta}-\frac{1}{2}\partial_\beta h^\alpha_{\;\alpha}=0$ . In this gauge, a plane wave has the form
\begin{equation}\label{1}
h_{ij}\sim\epsilon_{ij}e^{-i(\omega t-\textbf{k}\cdot \textbf{x})}+c.c.
\end{equation}
where $\epsilon_{ij}$ is the graviton polarization tensor. On the other hand, the Fermi normal coordinate allows us to write $h_{00}=-R_{0i0j}x^ix^j$. Thus, by taking into account the explicit form of the Riemann tensor as $R_{\mu\nu\alpha\beta}=\frac{1}{2}(\partial_\beta\partial_\mu h_{\alpha\nu}+\partial_\alpha\partial_\nu h_{\beta\mu}-\partial_\alpha\partial_\mu h_{\beta\nu}-\partial_\beta\partial_\nu h_{\alpha\mu})$ and with the help of (3) and (4) one obtains [5, 6]
\begin{equation}\label{1}
H_{int}=\frac{1}{4}m\omega^2h_{ij}x^ix^j=\frac{1}{2}h_{ij}T^{ij}.
\end{equation}
where $T_{ij}=\frac{1}{2}m\omega^2x_ix_j$ is the energy momentum tensor in a locally inertial frame. To be more specific and in terms of the interaction Lagrangian, we have
\begin{equation}\label{1}
L_{int}=\frac{1}{2}m\dot x_i\dot x_jh^{ij}+L'\xrightarrow{\textrm{locally inertial frame}}
L_{int}=-\frac{1}{4}m\omega^2x_ix_jh^{ij}.
\end{equation}
Note that the interaction between the bound potential and gravitational filed is absent in a locally inertial frame, i.e. $L'=0$.
\section{Transition Amplitude in Locally Inertial Frame}
The linearized Einstein-Hilbert action plus the interaction term in TT gauge, has the form
\begin{equation}\label{1}
S_{EH}+S_{int}=\frac{1}{16\pi G}\int d^4x \, h^{ij}D^{-1}_{ijkl}h^{kl}+\frac{1}{2}\int d^4x\, T_{ij}h^{ij}.
\end{equation}
where $T_{ij}=-\frac{1}{2}m\omega^2x_ix_j$. The classical equation of motion is
\begin{equation}\label{1}
\frac{\delta}{\delta h^{ij}}(S_{EH}+S_{int})=D^{-1}_{ijkl}h^{kl}+4\pi GT_{ij}=0.
\end{equation}
The Green function $D_{ijkl}(x-x')$ is defined via [18]
\begin{equation}\label{}
D^{-1}_{ijkl}D^{klmn}(x-x')=\frac{1}{2}(\delta_i^m\delta_j^n+\delta_i^n\delta_j^m)\delta(x-x').
\end{equation}
Therefore, with the aid of (9) one can invert (8) to write
\begin{equation}\label{}
h_{ij}(x)=-4\pi G\int d^4x'D_{ijkl}(x-x')T^{kl}(x').
\end{equation}
We are interested in transition amplitude of a particle moving from $\textbf{x}''$ at $t''=0$ to $\textbf{x}'$ at $t'=T$, assuming that no graviton is emitted during such a transition. The corresponding amplitude is [10]
{\setlength\arraycolsep{2pt}
\begin{eqnarray}\label{1}
\mathcal{A}(\textbf{x}',T;\textbf{x}'',0)&=&\langle0|\int_{\scriptsize{\textbf{x}^{\prime\prime}}} ^{\scriptsize{\textbf{x}^{\prime}}}D\textbf{x}\int Dh_{ij} \,e^{iS_0[\scriptsize\textbf{x}]
+iS_{int}[\scriptsize{\textbf{x}},h_{ij}]+iS_{EH}[h_{ij}]}|0\rangle.
\end{eqnarray}}
with $|0\rangle$ denoting the graviton vacuum state. Integration over the field $h_{ij}$ can be divided into the sum of classical and fluctuating fields. Integrating out the fluctuations and employing (7) and (10), yields the effective action $\Gamma$ through
\begin{equation}\label{1}
e^{i\Gamma}=\langle 0|\int D{h_{ij}}\exp i\bigg(\frac{1}{16\pi G}\int d^4x  h^{ij}D^{-1}_{ijkl}h^{kl}+\frac{1}{2}\int d^4xT_{ij}h^{ij}\bigg)|0\rangle,
\end{equation}
where
{\setlength\arraycolsep{2pt}
\begin{eqnarray}\label{2}
\Gamma&=&-\pi G\int d^4x_2\int d^4x_1\, T^{ij}({x}_2)D_{ijkl}({x_2}-x_1)T^{kl}({x}_1),\\\nonumber
&=&-\pi G\int\frac{d^3k}{(2\pi)^3}\int_0^T dt_2\int_0^T dt_1\,\tau^{ij}_{\scriptsize\textbf{k}}({t}_2)D_{ijkl,\scriptsize\textbf{k}}({t}_2-{t}_1)\tau^{kl}_{\scriptsize\textbf{k}}({t}_1).
\end{eqnarray}}
with the Fourier transform of the energy-momentum tensor as $\tau_{ij,\scriptsize\textbf{k}}=-\frac{1}{2}m\omega^2{x}_i{x}_je^{-i{\scriptsize\textbf{k}\cdot\textbf{x}}}$. The time-dependent part of Fourier transform of the Green function is
\begin{equation}\label{1}
D_{ijkl,\scriptsize\textbf{k}}(t-t')=\sum_{\lambda=1}^2\int_{-\infty}^\infty\frac{d\omega}{2\pi}\frac{e^{i\omega(t-t')}}{\omega^2-\textbf{k}^2+i\epsilon}
\epsilon^\lambda_{ij,\scriptsize\textbf{k}}\overline{\epsilon}^\lambda_{kl,\scriptsize\textbf{k}}.
\end{equation}
where $P_{ijkl,\scriptsize\textbf{k}}=\sum_{\lambda=1}^2\epsilon^\lambda_{ij,\scriptsize\textbf{k}}\overline{\epsilon}^\lambda_{kl,\scriptsize\textbf{k}}$ denotes the spin sum of polarization tensors. Consequently, the transition amplitude takes the form
{\setlength\arraycolsep{2pt}
\begin{eqnarray}\label{1}
\mathcal{A}(\textbf{x}',T;\textbf{x}'',0)&=&\int_{\scriptsize{\textbf{x}^{\prime\prime}}} ^{\scriptsize{\textbf{x}^{\prime}}}D\textbf{x} \,e^{iS_0[\scriptsize\textbf{x}]+i\Gamma[\scriptsize\textbf{x}]}.
\end{eqnarray}}
Performing the integral over $\omega$ yields ($ k=|\bf k|$)
\begin{equation}\label{1}
\int_{-\infty}^\infty\frac{d\omega}{2\pi}\frac{\omega^4e^{i\omega(t-t')}}{\omega^2-\textbf{k}^2+i\epsilon}
=\frac{k^3}{2i}e^{-ik|t-t'|},
\end{equation}
and the effective action modifies to
{\setlength\arraycolsep{2pt}
\begin{eqnarray}\label{2}
\Gamma=\frac{i}{2}\pi  G\int\frac{d^3k}{(2\pi)^3}k^3P_{ijkl,\scriptsize\textbf{k}}\int_0^T dt_2\int_0^T dt_1\,e^{-ik|t_2-t_1|}\xi^{ij}_{\scriptsize\textbf{k}}(t_2)\overline{\xi}^{kl}_{\scriptsize\textbf{k}}(t_1).
\end{eqnarray}}
where $\xi_{ij,\scriptsize\textbf{k}}=m{x}_i{x}_je^{-i{\scriptsize\textbf{k}\cdot\textbf{x}}}$.
\section{Graviton Emission Rate}
Now we consider the persistence amplitude of a given energy level, say $\phi_n$, of a bound system. We have [10, 15]
{\setlength\arraycolsep{2pt}
\begin{eqnarray}\label{1}
\mathcal{A}_{nn}(T)&=&\int d^3{x}'\int d^3{x}''\,\overline\phi_{n}(\textbf{x}')\mathcal{A}(\textbf{x}',T;\textbf{x}^{\prime\prime},0)
{\phi}_{n}(\textbf{x}^{\prime\prime}),\\\nonumber
&\approx&e^{i\Gamma_n}\int d^3x'\int d^3x''\,\overline\phi_n(\textbf{x}')
\langle\textbf{x}',T|\,\textbf{x}^{\prime\prime},0\rangle_0{\phi}_n(\textbf{x}''),
\end{eqnarray}}
where
{\setlength\arraycolsep{2pt}
\begin{eqnarray}\label{1}
\Gamma_n&=&\frac{i}{2}\pi  G\int\frac{d^3k}{(2\pi)^3k}P_{ijkl,\scriptsize\textbf{k}}\int_0^T dt_2\int_0^T dt_1e^{-ik|t_2-t_1|}\\\nonumber
&\times&\int_{\scriptsize\textbf{x}',\scriptsize\textbf{x}'',\scriptsize\textbf{x}_1,
\scriptsize\textbf{x}_2}{\phi}_{n}(\textbf{x}')\langle\textbf{x}',0|\,\textbf{x}_2,t_2\rangle_0 \xi^{ij}_{\scriptsize\textbf{k}}(t_2)\langle\textbf{x}_2,t_2|\,\textbf{x}_1,t_1\rangle_0 \overline{\xi}^{kl}_{\scriptsize\textbf{k}}(t_1) \langle\textbf{x}_1,t_1|\,\textbf{x}^{\prime\prime},0\rangle_0 \overline{\phi}_{n}(\textbf{x}'').
\end{eqnarray}}
Here, we have introduced the abbreviation
\begin{equation}\label{1}
\int_{\scriptsize\textbf{x}\ldots}\equiv\int d^3{x}\ldots
\end{equation}
and $\langle\textbf{x}',t'|\,\textbf{x}^{\prime\prime},t''\rangle_0$ stands for the the particle propagator
{\setlength\arraycolsep{2pt}
\begin{eqnarray}
\langle\textbf{x}',t'|\,\textbf{x}^{\prime\prime},t''\rangle_0&=&\int D\textbf{x} \,e^{iS_0[\scriptsize\textbf{x}]},\\\nonumber
&=&\sum_n{\phi}_n(\textbf{x}')\overline\phi_n(\textbf{x}'')e^{-iE_n(t'-t'')}.
\end{eqnarray}}
The particle's Hamiltonian supposed to be $H_0=\frac{\textbf{p}^2}{2m}+V(\textbf{x})$ with eigen-states $\phi_n$ satisfying $H_0\phi_n=E_n\phi_n$. The integral in second line of (18) has the form $e^{-iE_nT}$. Provided that $\frac{T}{\lambda}\ll 1$, which holds in $T\rightarrow\infty$ limit, one arrives at [15]
{\setlength\arraycolsep{2pt}
\begin{eqnarray}\label{1}
\int^T_0 dt_2\int^T_0 dt_1 e^{i(\lambda+i\epsilon) |t_2-t_1|}&=&2\int^T_0 dt_2\int^{t_2}_0 dt_1 e^{i(\lambda+i\epsilon) (t_2-t_1)}\qquad (t_1<t_2)\\\nonumber
&=&\frac{2iT}{\lambda+i\epsilon}.
\end{eqnarray}}
Note that we have excluded an irrelevant term arising on setting $t_2 = t_1$. Therefore, equation (19) becomes
{\setlength\arraycolsep{2pt}
\begin{eqnarray}\label{1}
\Gamma_n&=&-\pi TG\sum_{i=1}^2\int\frac{d^3k}{(2\pi)^3}k^3P_{ijkl,\textbf{k}}\sum_{n'} \frac{\langle n|\xi^{ij}_{\scriptsize\textbf{k}}|n'\rangle\langle n'|\xi^{kl}_{\scriptsize\textbf{k}}|n\rangle|}{E_{n}-E_{n'}-k+i\epsilon}.
\end{eqnarray}}
Now assuming the dipole approximation, i.e. $\xi^{ij}_{\scriptsize\textbf{k}}\approx mx_ix_j$ and employing [3]
\begin{equation}\label{2}
\int d\Omega_{\scriptsize\textbf{k}} P_{ijkl,\scriptsize\textbf{k}}=\frac{8\pi}{5}\Big(\frac{1}{2}\delta_{ik}\delta_{jl}
+\frac{1}{2}\delta_{il}\delta_{jk}-\frac{1}{3}\delta_{ij}\delta_{kl}\Big),
\end{equation}
one finds
\begin{equation}\label{}
\Gamma_n=-\frac{TG}{45\pi}\int_0^\infty{dk}k^5\sum_{n'} \frac{\langle n|Q_{ij}|n'\rangle\langle n'|Q^{ij}|n\rangle|}{E_{n}-E_{n'}-k+i\epsilon},
\end{equation}
where we have gained
\begin{equation}\label{1}
m^2x^ix^jx^kx^l\Big(\frac{1}{2}\delta_{ik}\delta_{jl}+\frac{1}{2}\delta_{il}\delta_{jk}-\frac{1}{3}\delta_{ij}\delta_{kl}\Big)=\frac{1}{9}Q_{ij}Q^{ij}.
\end{equation}
reminding that the quadruple moment is $Q_{ij}=m(3x_ix_j-x^kx_k\delta_{ij})$. Since $\sum_{n'(\neq n)}| \mathcal A_{n'n}|^2=1-|\mathcal A_{nn}|^2$, from
\begin{equation}\label{1}
\sum_{n'(\neq n)}| \mathcal A_{\scriptsize{n'n}}|^2=1-e^{-2\textrm{Im}\Gamma_n}\simeq2\textrm{Im}\Gamma _n,
\end{equation}
we obtain the graviton emission rate due to the process $|n,0\rangle\rightarrow|n',\textbf{k}\rangle$ as
\begin{equation}\label{1}
R_n=\frac{1}{T}\sum_{n'(\neq n)}| \mathcal A_{\scriptsize{n'n}}|^2=\frac{2}{T}\textrm{Im}\Gamma_n.
\end{equation}
provided that $\textrm{Im}\Gamma_n\ll 1$. The imaginary part of the (25) can be obtained by applying
\begin{equation}\label{1}
\textrm{Im}\frac{f(x)}{x-x_0-i\epsilon}=\pi f(x)\delta(x-x_0).
\end{equation}
So, from one (28) one finds the graviton emission rate as
{\setlength\arraycolsep{2pt}
\begin{eqnarray}\label{1}
 R_n=\frac{2}{45}G\sum_{n'} \omega_{nn'}^5\langle n|Q_{ij}|n'\rangle\langle n'|Q^{ij}|n\rangle.
\end{eqnarray}}
On multiplying the above expression by the emitted graviton energy $\omega_{nn'}$, we obtain the intensity of radiation due to the decay of eigen-state $|n\rangle$
\begin{eqnarray}\label{1}
-\frac{dE_n}{dt}=\frac{2}{45}G\sum_{n'} \omega_{n'n}^6\langle n|Q_{ij}|n'\rangle\langle n'|Q^{ij}|n\rangle,\qquad (n'<n).
\end{eqnarray}
This result for the radiation intensity could be achieved in TT gauge, by taking into account the the interaction between the potential and gravitational field [10]. It involves to start from (7), supposing the interaction Lagrangian as the left-hand side of (6). However, as equation (31) entails, the physics is independent of any reference frame and moving on a locally inertial frame provides us with more economical way of carrying out the calculations.


\begin{thebibliography}{99}
\bibitem{1} S. Weinberg,
\emph{Gravitation and Cosmology}, John Wiley \& Sons, New York (1972), Chap. 10.
\bibitem{1} C. Kiefer,
\emph{Quantum Gravity}, Clarendon Press, Oxford (2004), Chap. 2.
\bibitem{1} M. D. Scadron,
\emph{Advanced Quantum Theory}, 3rd Edition, World Scinetific, London (2007), Chap. 14.
\bibitem{2}{L. Smolin, Gen. Rel. Grav. \bf 17} (1985) 417-437 .
\bibitem{2}{T. Rothman, S. Boughn,  Found. Phys. \bf 36} (2006) 1801-1825.
\bibitem{2}{S. Boughn, T. Rothman, Class. Quant. Grav. \bf 23} (2006) 5839-5852.
\bibitem{2}{A. D. Speliotopoulos, Phys.Rev. \bf D51} (1995) 1701-1709.
\bibitem{2}{ G. Schafer, H. Dehnen, J. Phys. A: Math. Gen. \bf13} (1980) 2703-2722.
\bibitem{2}{ N. Grafe, H. Dehnen, Int. J. Theor. Phys. \bf15} (1976) 393-409.
\bibitem{2}{ G. Schafer, H. Dehnen, J. Phys. A: Math. Gen. \bf14} (1981) 2359-2365.
\bibitem{2}{ G. Schafer, H. Dehnen, Phys. Rev. \bf D23} (1981) 2129-2136.
\bibitem{2}{ F. Romero, H. Dehnen, Astrophysics and Space Science \bf 89} (1983) 115-125.
\bibitem{2}{L. Halpern and R. Desbrandes, Ann. Inst. Henri Poncare \bf A11} (1969) 309-329.
\bibitem{2}{L. P. Grishchuk, Phys. Rev. \bf D45} (1992) 2601-2608.
\bibitem{4} B. R. Holstein,
\emph{Topics in Advance Quantum Mechanics}, Addison-Wesley (1992), Chap. 4.
\bibitem{5} R. P. Feynamn, A. R. Hibbs
\emph{Quantum Mechanics and Path Integrals}, McGraw-Hill (1965), Chap. 9.
\bibitem{2}{B. S. DeWitt, Rev. Mod. Phys. \bf 29} (1957) 377-397.
\bibitem{2}{G. t'Hooft, M. Veltman, Ann. Inst. Henri Poncare \bf A20} (1974) 69-94.
\end{thebibliography}
\end{document}